\documentclass[twocolumn,showpacs,superscriptaddress,prb]{revtex4}

\usepackage{graphicx}
\usepackage{amsmath}
\usepackage{ifpdf}
    \ifpdf
				\usepackage[pdftex]{graphicx}
		\else
					\usepackage{graphicx}
    \fi

\begin{document}
\title{Spatially-indirect Exciton Condensate Phases in Double Bilayer Graphene}

\author{Jung-Jung Su}
\email{jungjsu@nctu.edu.tw}
\affiliation{Department of Electrophysics, National Chiao Tung University, Hsinchu 300, Taiwan}

\author{Allan H. MacDonald}
\email{macd@physics.utexas.edu}
\affiliation{Department of Physics, University of Texas at Austin, Austin, TX 78712, USA}

\date{\today}

\begin{abstract} 
We present a theory of spatially indirect exciton condensate states in systems composed of 
a pair of electrically isolated Bernal graphene bilayers.  
The ground state phase diagram in a two-dimensional displacement-field/inter-bilayer-bias
space includes layer-polarized semiconductors, spin-density-wave states,
exciton condensates, and states with mixed excitonic and spin order.
We find that two different condensate states, distinguished by a chirality index,
are stable under different electrical control conditions.
\end{abstract}


\maketitle

\section{Introduction}

\begin{figure}
\includegraphics[width=.8\linewidth]{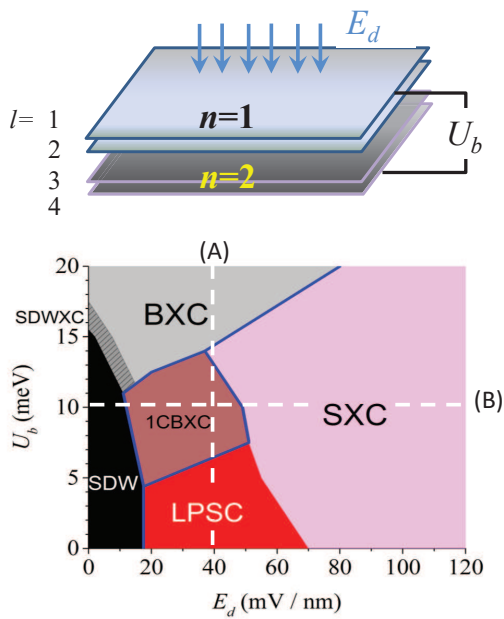}
\caption{Upper panel: Schematic illustration of the experimental system 
modeled in this paper. 
An external electric field $E_d$ is applied simultaneously to two
bilayers (bilayer $n=1$ with layers $l=1,2$ and 
bilayer $n=2$ with layers $l=3,4$) by external gates.
Separately an electrical bias potential
$U_b$ is applied between the two bilayers. 
In this paper we assume that the leakage current between bilayers is negligible.
Lower panel: Mean-field theory phase diagram of electrically neutral double-bilayer graphene as a function of  
$U_b$ and $E_d$.  In this paper we have not allowed valley-symmetry to be broken.  Decoupled bilayers 
then have a first order transition between a spin-density-wave (SDW) state at small displacement fields and a 
layer polarized semiconductor (LPSC) state with no spontaneously broken 
symmetries at large displacement fields.  In double-bilayers 
the LPSC is unstable at large displacement fields $E_d$
to an exciton condensate (SXC) with coherence mainly between the 
adjacent single layers $l=2$ and $l=3$. 
Excitons can also be induced electrically by applying a bias voltage $U_b$ between bilayers.
For large $U_b$ the SDW state is unstable to an exciton condensate (BXC) with coherence 
mainly between the conduction band of bilayer $n=2$ and the 
valence band of bilayer $n=1$.  Within each bilayer  
the pattern of interlayer coherence is determined by single-particle physics. 
The differences between the order parameters of the SXC and BXC states, and those of other states 
with both excitonic and spin order that occur at intermediate 
values of $U_b$ and $E_d$ are discussed at length in the main text.
The phase boundaries marked by solid blue lines are first order, and the remaining phase 
transitions are continuous.  We explain the physics of the various phase transitions
below by closely examining competing states along lines (A) and (B).  
This phase diagram was constructed for the case of
inter-bilayer separation $t_{\rm hBN}=0.6$nm; $t_{\rm hBN}$-dependence is discussed at length, and 
the corresponding phase diagram at $t_{\rm hBN}=1.2$nm is 
presented in the penultimate section of this paper.  
}
\label{fig:Setup&PD}
\end{figure}

Bernal stacked bilayer graphene is\cite{McCannFalko2006} a two-dimensional semiconductor with an 
electrically tunable band gap that can be as large as $\sim 250$ meV \cite{Rotenberg2006, McCann2006, Oostinga_NatMat2007,Castro_PRL2007,Min2007,Nilsson2007,Zhang2009}
when external gates are used to apply a large displacement field perpendicular to the graphene planes.  The optical spectrum
of bilayer graphene features\cite{Hoon2008,Mak_PRL2009,Louie2010} strong and atypical excitonic features.
The exciton binding energy increases along with the band gap in strong displacement fields,
and can be as large as tens of meVs.  When  two graphene bilayers are placed in close proximity,
spatially-indirect excitons typically exist as elementary excitations, but  
can also be present in the ground state under relatively easily established gating conditions.
Our main interest in this paper is in constructing a phase diagram for double bilayer graphene systems 
in the absence of a magnetic field that incorporates the possibility of equilibrium condensation of spatially indirect excitons. 
 
Spatially indirect exciton condensates have a surprising and 
fundamentally interesting suite of anomalous transport properties that  previously 
have been studied\cite{Eisenstein2004,EisensteinReview}
only in semiconductor-quantum-well bilayers and only in the quantum Hall regime.  They 
are counterflow superfluids and exhibit, among other properties,
spontaneous phase coherence between electrically isolated subsystems.
These condensates can be manipulated by external electrical contacts\cite{Su2008,Su2010,Pesin2011,Hsu_SR2015(Su)} via 
an excitonic generalization of Andreev scattering.  Excitonic superflow is manifested most explicitly
in a variety of transport experiments,\cite{vonKrecent,Eisensteinrecent}  in which
the electrically isolated two-dimensional electron systems are contacted independently.  

In the quantum Hall regime, 
the dissipationless flow of quasiparticle 
charges in chiral edge channels plays an essential role in 
determining how excitonic and charged quasiparticle 
currents interact with external bias voltages. 
In addition it appears\cite{Inti2014} that for currently available
quantum Hall bilayer superfluids, disorder also
has a large impact on certain quantitative aspects of the observed phenomena.  
If spatially indirect exciton condensation could be achieved in the absence 
of a magnetic field, the absence of edge states would introduce new and fundamentally interesting differences.
In particular, it seems likely that disorder could play a less essential role, facilitating 
more quantitative comparisons between theory and experiment.  Double bilayer graphene is an attractive 
system for efforts to achieve spatially indirect exciton condensation because the energy gaps in 
each bilayer are relatively small making it easier to electrically induce equilibrium exciton populations,
because the properties of each two-dimensional electrical system can be tuned electrically
by using gates to vary their internal displacement fields, and because great progress has been
achieved experimentally in the flexible construction of multilayer graphene systems with very weak disorder.  

The physics of bilayer graphene is rich even when only one bilayer is present. 
The ground state at the charge neutrality point in the absence of an external displacement field is an interesting spin-density-wave 
state with opposite spin-orientations on opposite layers and a very small staggered moment per atom.\cite{Maher_NatPhys2013,LauCommentary}  
A first order phase transition\cite{CommentAHE} occurs near displacement field 
$E_d \sim 15$ mV/nm from this spin-density-wave state to 
a layer-polarized two-dimensional semiconductor state without any broken symmetries.  
At large displacement fields, the semiconductor state 
gap can be viewed as originating from an avoided crossing between the conduction band of the low electric potential
layer and the valence band of the high electric potential layer.  The size of the gap is then limited by the 
strength of interlayer tunneling in the bilayer.  
The excitons of this semiconductor are 
unusual\cite{Hoon2008,Louie2010,BilayerGW, Berman2012,Franz2008,Pikulin_NatComm2016} because of the 
Berry phase properties graphene's two-dimensional Dirac model states and are in this 
sense similar to the excitons of a topological insulator \cite{Franz2009}.  The properties of optically excited populations of 
bilayer graphene excitons which have thermalized and condensed have been
studied in previous\cite{Berman2012} theoretical work.  Our interest however is in excitons which 
are present in thermal equilibrium, and this can be achieved only for spatially indirect excitons 
and only when two bilayer graphene systems are present and separated by an insulating barrier.\cite{Li_arxiv2016(Dean_doublebilayer), Liu(Kim), Neilson2013,Zarenia_SciRep2014,Tutuc2016}

Spatially indirect excitons generally have smaller binding energies than spatially direct excitons because of the 
increased separation between electrons and holes.  However their excitation energies can be tuned  
electrically by using a gate-controlled displacement field to adjust the relative band line-up 
of the two bilayers.    
For a fixed spatial separation between bilayers a displacement field qualitatively 
alters the band structure of each bilayer, adjusting the band gaps and also the 
exciton energies.  
The sensitivity of isolated bilayers to displacement fields plays an essential role 
in the physics we explore below.  
Previous theoretical work\cite{Neilson2013,Zarenia_SciRep2014} has highlighted 
the potential of double bilayer graphene as a spatially indirect exciton system, 
but assumed simplified parabolic band dispersions for conduction and valence bands 
and did not explore the consequences either of this displacement-field sensitivity or 
of broken symmetry states in isolated bilayers.  
  
The properties of double bilayer systems can be adjusted electrically by using gates to apply a 
displacement field or by applying a bias voltage between bilayers, 
as illustrated in Fig.[~\ref{fig:Setup&PD}].  
When tunneling between bilayers can be neglected, bias voltages have the advantage that they alter 
spatially indirect exciton energies without changing the properties of 
the isolated bilayers.   In this paper we do not 
account for inter-bilayer tunneling, {but} 
concentrating instead on establishing the phase diagram
of the negligible tunneling limit.  
The strength of inter-bilayer tunneling declines exponentially
with the number of layers of intervening dielectric, but is sensitive\cite{Bistritzer_PRB2010} to disorder and to the 
relative orientations of both graphene bilayers and the hBN barrier layers.  
An important difference between graphene monolayers and bilayers is that the latter are semiconductors with a gap \cite{LauExperiment} between conduction and valence bands, whereas the former are gapless.  We are interested here in chemical potential differences between bilayers that are smaller than the gap by an amount close to the exciton binding energy, 
an interval of bias voltage over which direct the interband tunneling rate vanishes in the temperature $ T \to 0$ limit
when disorder is absent.  {We therefore expect that in the regime of interest the tunneling-assisted charge equilibration times between bilayer will be 
substantially larger than those between monolayers ($\sim 10^{-8} s$ for one-layer-thick hBN barriers), and probably dominated by 
electro-luminescence processes.}   We also expect that the quasi-equilibrium approximation we
employ, which will be reliable provided that the charge equilibration times exceed 
thermalization times, which are often smaller than a ps,\cite{Mihnev_NatComm2015,Song_JPhysC2015,Bistritzer_PRL2009},
can be applicable down to the smallest barrier thicknesses in highly perfect samples. 
The physics of systems in which the quasi-equilibrium approximation 
is not valid lies beyond the scope of the present work, but is interesting, and related to
phenomena that have been studied in polariton condensate systems.\cite{PolaritonRefs}

\setlength{\tabcolsep}{10pt}
\begin{table}[t]
\begin{tabular}{ l  l   }
\hline
\hline
  SDW & Spin Density Wave \\
  LPSC & Layer-Polarized SemiConductor \\
  SXC & Single-layer Exciton Condensate \\
	BXC & Bilayer Exciton Condensate \\
	SDWXC & Spin-Density-Wave Exciton Condensate \\
	1CBXC & 1-component Bilayer Exciton Condensate\\
\hline
\hline
\end{tabular}
\caption{Acronyms used for distinct electronic states in the main text.  {The SDW and 
SDWXC states break time-reversal symmetry.  The SXC, BXC, SDWXC, and 1CBXC states break
the separate particle-number conservation symmetry of the individual bilayers.}}
\label{Tab:acronym}
\end{table}

Exciton condensation occurs when  
the excitation energy needed to create a spatially indirect exciton in 
a bilayer graphene system has been electrically adjusted to a negative value, leading to  
a finite density population of excitons in equilibrium.  Our results for the 
phase diagram of double-bilayer graphene are  summarized in Fig.[~\ref{fig:Setup&PD}]
which illustrates how the double-bilayer state depends on the displacement field $E_{d}$ and on the  
electrical bias energy $U_{b}$.  
The phase diagram in Fig.[~\ref{fig:Setup&PD}] was calculated for bilayer separation $t_{\rm hBN} = {0.3} {\rm nm}$, corresponding to a single layer of hexagonal boron-nitride (hBN) between the bilayers.
The corresponding phase diagram for a bilayer separation of $t_{\rm hBN} = { 0.9} {\rm nm}$, corresponding to 
the case of three intervening hBN layers, has been calculated as well. 
We choose to discuss the $t_{\rm hBN} = {0.3} {\rm nm}$ case first, because the phase diagram is 
richest at small layer separations.  In addition the phase diagram at larger 
values of $t$ can be very accurately extrapolated from the $t_{\rm hBN} = {0.3} {\rm nm}$ results using a 
procedure we explain later.   

{We describe the many-exciton state using mean-field theory.} We find that for small $U_b$ the two bilayers are uncorrelated, and that they   
have a first order transition between a spin-density-wave (SDW) state at small displacement fields $E_d$ and a 
layer-polarized semiconductor (LPSC) state at larger displacement fields. The LPSC is unstable at still larger $E_d$ 
to the SXC state in which coherence is established mainly between the 
adjacent single graphene layers labelled $l=2$ and $l=3$. 
For large $U_b$ and small $E_{d}$ the SDW state is unstable to the bilayer exciton condensate (BXC) state 
in which coherence is established mainly between the conduction band of bilayer $n=2$ and the 
valence band of bilayer $n=1$. 
Among other interesting features that appear in this phase diagram, 
we find that exciton condensation is sometimes combined with spin-density-wave order, which breaks time-reversal symmetry,
and sometimes not, and that excitons condense into different states at large $E_{d}$ than at large $U_{b}$. 
The phase diagrams at larger $t_{\rm hBN}$s are discussed in the later sections. 
The greatest change is that single-layer exciton condensate (SXC) states occurs at smaller values of $E_d$
because the external potential difference between the bilayers at a given $E_d$ increases.
The reduction in exciton binding energies at larger $t_{\rm hBN}$s also plays a role. 

Note that the mean-field theory does not account for quantum fluctuations in the many-body ground state which are evidenced, for example, by finite drag resistivities in the absence of interlayer coherence.  When quantum fluctuations are included, states with interlayer phase coherence can lose their order and form Fermi liquid states, or excitons can pair to form biexcitons.  All of these possibilities are discussed at greater length later.  

Our paper is organized as follows.  In Section II we provide some technical details on the mean-field calculation we perform for 
the four layer graphene system of interest, the double-bilayer.
Section III  summarize the mean-field theory description of the 
SDW to LPSC state phase transition in an isolated bilayer, showing that reasonable agreement with 
experiment follows from a physically sensible approximate treatment of screening that we employ for
all subsequent calculations.  
In Section IV we describe in detail how the quasiparticle energy bands and wave functions evolve as $E_{d}$ is 
increased at $U_{b}=0$.  Along this line in the phase diagram, reduction in spatially indirect exciton energy 
with increasing $E_{d}$, is accompanied by increasing layer polarization within the individual bilayers.
For bilayers separated by a single-layer of hBN, we find that the SDW to LPSC phase transition
occurs before the indirect exciton energy vanishes and interlayer phase coherence appears.  
By the time the phase coherent condensate state appears, the individual bilayers are already strongly polarized and 
the condensate is dominated by coherence between the most closely spaced layers.  In Section V  we describe in 
detail how the quasiparticle energy bands and wave functions evolve as $U_{b}$ is varied at $E_{d}=0$.
In this case the individual layers have a small gap associated with spin-density-wave order.  Condensation 
then occurs first as an instability of a state with broken time-reversal symmetry. 
Coherence is strongest between layers 1 and 3 for one spin orientation and between layers 2 and 4 for the other. 
{Upon increasing $U_b$ further, the condensate evolves into the BXC state which is spin-rotationally invariant and allows for charge transfer between the bilayers.}
In Section VI, we discuss lines (A) and (B) lines in the phase diagram of Fig.~\ref{fig:Setup&PD}).
When $E_d$ and $U_b$ are present, a series of intermediate states can occur along lines which 
cross between weakly-correlated bilayer and either SXC (large $E_d$) or BXC (large $U_b$) states.
We also present a phase diagram calculated by applying the same considerations to 
a model with a larger bilayer separation, $t_{\rm hBN}={0.9}$ nm, 
and explain how it is related to the $t_{\rm hBN} = {0.3}$ nm phase diagram. 
Finally in Section VII we comment on the limitations of mean-field theory, and speculate on the experimental implications of this study.

\section{Mean-Field Theory of Exciton Condensate and Spin-Density-Wave States in Double Bilayer Graphene}

The mean-field theory calculations preformed here
neglecting the possibility of broken valley symmetry, but allow the Hamiltonian's spin-rotational-invariance to 
be broken.  We therefore study a 16-band model with $\pi$ orbitals of both spins on both sublattices 
of four honeycomb lattice layers.  
The full mean-field Hamiltonian is $H = H_B+H_{int}$ where $H_B$ is the single-particle band 
Hamiltonian and $H_{int}$ describes the Coulomb interaction contribution.
The band Hamiltonian can be written down most concisely as 
\begin{equation} 
H_B ({\bf k}) = (H_{BL} ({\bf k}) \otimes \sigma_0 + (U_b+eE_d\,(t_{\rm hBN}+2 d))/2\otimes \sigma_z )\otimes s_0
\end{equation}
where $H_{BL}$ is the band Hamiltonian of an isolated 
$AB$-stacked bilayer, $\boldsymbol \sigma$ is a Pauli matrix vector that
acts on the $which-bilayer$ psuedospin, and $\boldsymbol s$ is Pauli matrix that
acts on the real spin degree-of-freedom. 
Here we define $d$ as the separation between layers within a graphene bilayer and $t_{hBN}$ as the increase in the separation between graphene bilayers when they are separated by hBN layers, i.e. it is approximately 
equal to $d$ times the number of hBN layers that are present.  

In the  $(1A, 1B, 2A, 2B)$ sublattice representation, 
the isolated bilayer Hamiltonian is:\cite{McCannFalko2006} 
\begin{eqnarray} \label{HBL}
&& H_{BL} ({\bf k}) \nonumber \\
&& = \left( \begin{array}{cccc}
eE_d\,d/2 & \hbar v k \,e^{-i \psi_{\bf k}} & 0 & 0 \\
\hbar v k \,e^{i \psi_{\bf k}} & eE_d\,d/2 & \gamma_1 & 0\\
0 & \gamma_1 & -eE_d\,d/2 & \hbar v k \,e^{-i \psi_{\bf k}} \\
0 & 0 & \hbar v k \,e^{i \psi_{\bf k}} & -eE_d\,d/2\\
\end{array} \right) 
\end{eqnarray}
Here $v$ is the bare Dirac velocity of an isolated graphene layer,
$\gamma_1$ is the interlayer hopping parameter, and $\psi_{\bf k} \equiv \tan^{-1} (k_y/k_x)$.
 
We will see that the momentum-orientation-dependence of the inter-sublattice hopping term within each 
graphene layer, 
which is famously
responsible for Berry phase features in the 
electronic structure of all single and multi-layer graphene systems, also plays an important role 
in determining the double-bilayer phase diagram.     

In mean-field theory, Coulomb interactions give rise to self-consistently determined 
Hartree and exchange self-energies.  
We separate these contributions, writing $H_{int}=H_H+H_X$ where $H_H$ is the Hartree contribution and $H_X$ accounts for 
exchange.  
We use a representation of site-dependent $\pi$-band orbitals and  
label our 16 bands by the compound index $b \equiv \{l, x\}$ where $l=1 \ldots 4$ is the layer index and $x=A, B$ is the 
sublattice index within a layer.  With this notation the Hartree term in the mean-field Hamiltonian is 
\begin{equation} 
\label{HH}
\langle b' \,|\, H_H ({\bf k}) \,| b \rangle
=  - g  \, \delta_{b',b}  \frac{2 \pi e^2 }{\epsilon}  \,  \sum_{b''} d_{b,b''} \, n_{b''} .
\end{equation}
Here the factor $g=2$ accounts for the two-fold valley degeneracy, the dielectric constant $\epsilon$ is that of the 
embedding material, $ d_{bb'}$ is the distance between the layers associated with labels
$b$ and $b'$, and $n_{b}$ is the total carrier density projected onto band $b$. 

The exchange contribution to the mean-field Hamiltonian is responsible for exciton condensation and therefore
plays the most essential role in the physics described below.  
\begin{equation}
\label{HX}
\langle b|\, H_X ({\bf k}) \,| b'\rangle 
= -  \int \frac{d^2{\bf k'}}{(2 \pi)^2} \, V_{bb'}  (|{\bf k}-{\bf k}'|) 
\, \langle b|\, \tilde{\rho} ({\bf k}')\,| b'\rangle ,
\end{equation}
 where 
$V_{bb'}  (|{\bf q}|) = {2\pi e^2} \, e^{-q \, d_{bb'}} /{\epsilon q}$ is the two-dimensional Coulomb interaction
between bands $b$ and $b'$.  In evaluating Eq.(~\ref{HX}) we employ the regularized density matrix 
$\tilde{\rho} \equiv \rho-\rho_0$ where $\rho_0$ is the density matrix for isolated layers with 
full valence bands and empty conduction bands.  In doing so, we take the view that the
Dirac velocity parameter of an 
isolated layer already accounts for exchange interactions with
the bare valence band states.
Note that coherence between bands gives rise to an interaction-induced 
interband hopping term in the mean-field Hamiltonian.

The densities and density-matrices in Eqs. ~\ref{HH} and ~\ref{HX} must be 
determined self-consistently.  The off-diagonal terms in the density matrices capture 
the coherence between sites presents in the wavefunctions of occupied quasiparticle states.
Because of the dependence of hopping within each graphene layer on $\psi_{\bf k} \equiv \tan^{-1} (k_y/k_x)$,
the exchange contribution to the Hamiltonian is dependent on both momentum magnitude and orientation.  
Fortunately, the momentum orientation of the relative phases\cite{Hongkiprb2015}
of quasiparticle projections onto different bands has a very simple form in multilayer graphene:
\begin{equation}
\langle b|\, \tilde{\rho} ({\bf k}) \,|b' \rangle = f_{bb'}(k) \exp [-i (J_b-J_{b'}) \psi_{\bf k}],
\end{equation} 
where $f_{bb'}(k)$ captures the density-matrix dependence on $k$ and  
$J_b$ is a band-dependent chirality index which will be discussed at greater length below.  
With this notation, the exchange contributions to the mean-field Hamiltonian becomes: 
\begin{eqnarray}
\langle b| H_X ({\bf k}) | b' \rangle 
=  - e^{-i (J_b-J_{b'}) \psi_{\bf k}}  \int d k' 
u_{bb'}(k,k') f_{bb'}(k') \nonumber
\end{eqnarray}
The interaction factor $u_{b,b'}(k,k')$ is determined by an angular integral which can be evaluated once and for all, 
and used throughout the self-consistent iteration process: 
\begin{eqnarray}
u_{b,b'}(k,k') \equiv \frac{k'}{(2\pi)^2} \int d\theta \, V_{bb'} (q(k, k', \theta)) \, e^{-i (J_b-J_{b'}) \,\theta}. \nonumber
\end{eqnarray}
where $q(k, k', \theta)= (k^2+k'^2+2 k k' \cos(\theta))^{1/2}$.  
Note that exchange interactions are stronger between bands with nearby layers 
and more similar chirality indices.  

When interactions are neglected, the band-dependent chirality indices $J_b$ 
in an electrically coupled multilayer graphene system are determined by the stacking sequence.\cite{MinMacDonald} 
The index can be defined as: $J_b \equiv {\rm Arg} [\langle \phi \,({\bf k}) | b, {\bf k} \rangle]/ \psi_{\bf k}$, 
where $| \phi \,({\bf k}) \rangle$ is a site representation Bloch state.
Only the relative chirality indices between bands are gauge-invariant.
For a Bernal bilayer the chirality indices can be read off Eq.~\ref{HBL} by observing that the band eigenstates have 
momentum orientation dependence of the form: 
$|\phi \rangle = (c_{1A}, c_{1B} e^{i \psi_{\bf k} }, c_{2A}e ^{i \psi_{\bf k} }, c_{2B} e^ {i 2 \psi_{\bf k} })$, 
corresponding to chirality indices $J_{BL}=(0, 1, 1, 2)$ for the four sites $(1A, 1B, 2A, 2B)$.  

In a spatially indirect exciton condensate state coherence is spontaneously established between the Coulomb-coupled 
but electrically isolated bilayers of a double bilayer system. 
Because the band energy cost of altering band-chirality differences within either bilayer is prohibitive,
the band chirality indices in double bilayers are of the form 
$J_b = J_{BL}+ \delta_{n,2} J_{X}$.  ( Note that we are free to  choose a gauge 
with $J_{X}=0$ for bilayer $n=1$.)  $J_{X}$ 
can be viewed as the angular momentum of the spatially indirect excitons 
that form, and is an integer-valued chirality index that distinguishes different excitonic states.
Below we describe the properties of quasiparticle states in one graphene valley only, say valley K. 
Because we assume that the two valleys are related by time-reversal symmetry, the
chirality index of valley K' is understood to be opposite to that of valley K.  
 
With these conventions the $J_{X}=-2$ state is one in which the chirality difference between sites $1A$ and $4B$ is zero.
The self-consistent state with this choice for $J_{X}$ therefore has strong 
coherence between these two sites.  Similarly, the $J_{X}=0$ state has strong coherence
between equivalent sites in the two-bilayers.  We therefore refer to $J_X=0$ 
exciton condensates as bilayer exciton condensates (BXCs).  For $J_{X}=2$, the chirality
indices are identical for the $2B$ and $3A$ sies.  We refer to the $J_{X} = 2$ states 
as single layer exciton condensates (SXC).   
For each many-body state, we calculate total energy density by using
\begin{equation}  
\label{eq:totalenergy}
E_{tot}/A =\langle H_B \rangle/A + (\langle H_H \rangle+\langle H_X \rangle)/2A,
\end{equation} 
where $A$ is the area of the system. The factor of 2 in Eq.~\ref{eq:totalenergy} 
corrects for the double-counting of interactions in the mean-field state.
As we explain below the $J_X=0$ BXC and $J_X=2$ SXC states compete for the ground state, with the preferred state 
determined mainly by the influence of the displacement field $E_d$ on layer polarization within the isolated bilayers. 
The $J_{X}=-2$ states are never ground states.

It is important to recognize that there are small\cite{accuratebilayer}
but non-zero corrections to the minimal $\pi$-band model
Hamiltonian we have adopted in Eq.~\ref{HBL}, for example the corrections
responsible for trigonal warping of constant energy surfaces.  
When these terms are included, the quasiparticle bands do not have definite chirality indices 
even at the single-particle level.  Because these terms are small, however, we do not 
expect that they will materially influence the double bilayer phase diagram.

\begin{figure}
\includegraphics[width=.8\linewidth]{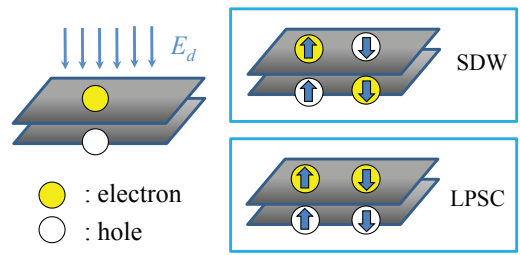}
\caption{Schematic illustration of the isolated bilayers states.  The competition
between spin-density-wave (SDW) and layer-polarized semiconductor (LPSC) states can be 
tuned experimentally by varying the displacement field $E_d$.  The 
SDW state has broken time reversal symmetry, whereas the 
layer-polarized-semiconductor (LPSC) state has no broken symmetries.
Both states have energy gaps.  The yellow and white circles represent electrons and holes respectively. 
The arrows in the circles represent spin-orientations.  In the SDW state, opposite spins have 
opposite layer polarization, whereas in the LPSC state, the layer polarization is 
not spin-dependent.}
\label{fig:Cartoon_1BL}
\end{figure} 

\section{Influence of a Displacement Field $E_{d}$ on Bilayer Graphene} 

In this section we discuss the application of mean-field theory to isolated bilayers subject to an applied displacement field 
$E_d$.  This digression is necessary partly because excitonic states emerge in many cases as 
instabilities of single-bilayer states that are on their own non-trivial, and partly as a reality check 
in which our approach is applied to a case in which extensive experimental data are already available.   
All calculations in this paper were performed using $e^2/2\pi \epsilon = 50$ meV nm, corresponding to $\epsilon \sim 4$,
and reducing interaction strengths by a further factor of $C_s = 0.8$ to account for 
additional screening effects. This value for $C_s$ was chosen phenomenologically to adjust the displacement field 
at which the SDW to LPSC transition (see below) occurs to $E_d \sim 20$ mV/nm, the value found 
experimentally by Velasco {\it et al.}.\cite{LauExperiment}

\begin{figure}
\includegraphics[width=.8\linewidth]{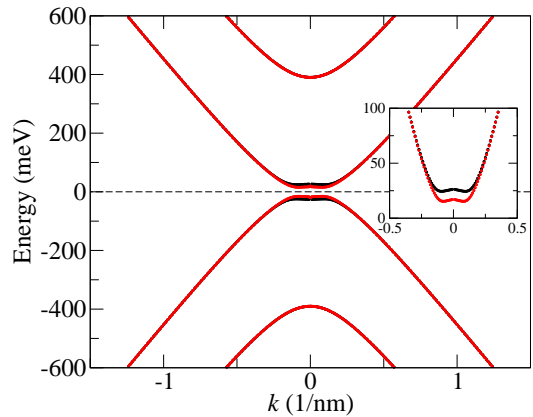}
\caption{Typical quasiparticle band dispersions for bilayer graphene at $E_d = 0$.
The black and red squares illustrate the quasiparticle energy 
dispersion of the SDW and LPSC states respectively. 
The bands of the two states are similar and both have gaps.  (See inset.)
Interactions dominate at small $k$ because of bilayer graphene's quadratic
crossing between valence and conduction bands.  The quasiparticle dispersion at large k 
is not strongly influenced by electron-electron interactions.}
\label{fig:E=0.0_1BL}
\end{figure}   

It is well established\cite{BilayerBrokenSymmetry,LauExperiment,Mayorov_Science2011,Freitag_PRL2012,Veligura_PRB2012,Bao_PNAS2012, Zhang_PRL2011, Lemonik_PRB2010,Castro_PRL2008,Zhang_PRL2012,Gorbar_PRB2010} that  at zero magnetic field 
the stable many-body states in an isolated bilayer graphene system are the 
spin-density-wave (SDW) and the layer-polarized-semiconductor (LPSC) states 
illustrated in Fig.~\ref{fig:Cartoon_1BL}.  In an SDW state, opposite layers have opposite spin-polarizations, 
although each layer is neutral only when $E_d=0$.  The SDW state breaks time reversal symmetry.
The microscopic character of this state is quite distinct from that of a Heisenberg model 
system on a honeycomb lattice, as signaled by the fact that the staggered moment  per atom is 
very small compared to one Bohr magneton per atom.  The scale of the spin polarization is 
in fact set by the strength of the interlayer tunneling amplitude,
which increases the masses of the states at valence and conduction band edges.  
Because the SDW state has an unfavorable layer polarization for one spin-orientation, it 
becomes unstable at large $E_d$.  The LPSC state, which is the ground state at large $E_d$ has no broken symmetry 
when $E_d \ne 0$.  Both states exhibit gaps (even for zero $E_d$) when electron-electron interactions are included,
as seen in Fig.~\ref{fig:E=0.0_1BL}.  

\begin{figure}[htb]
\includegraphics[width=.9\linewidth]{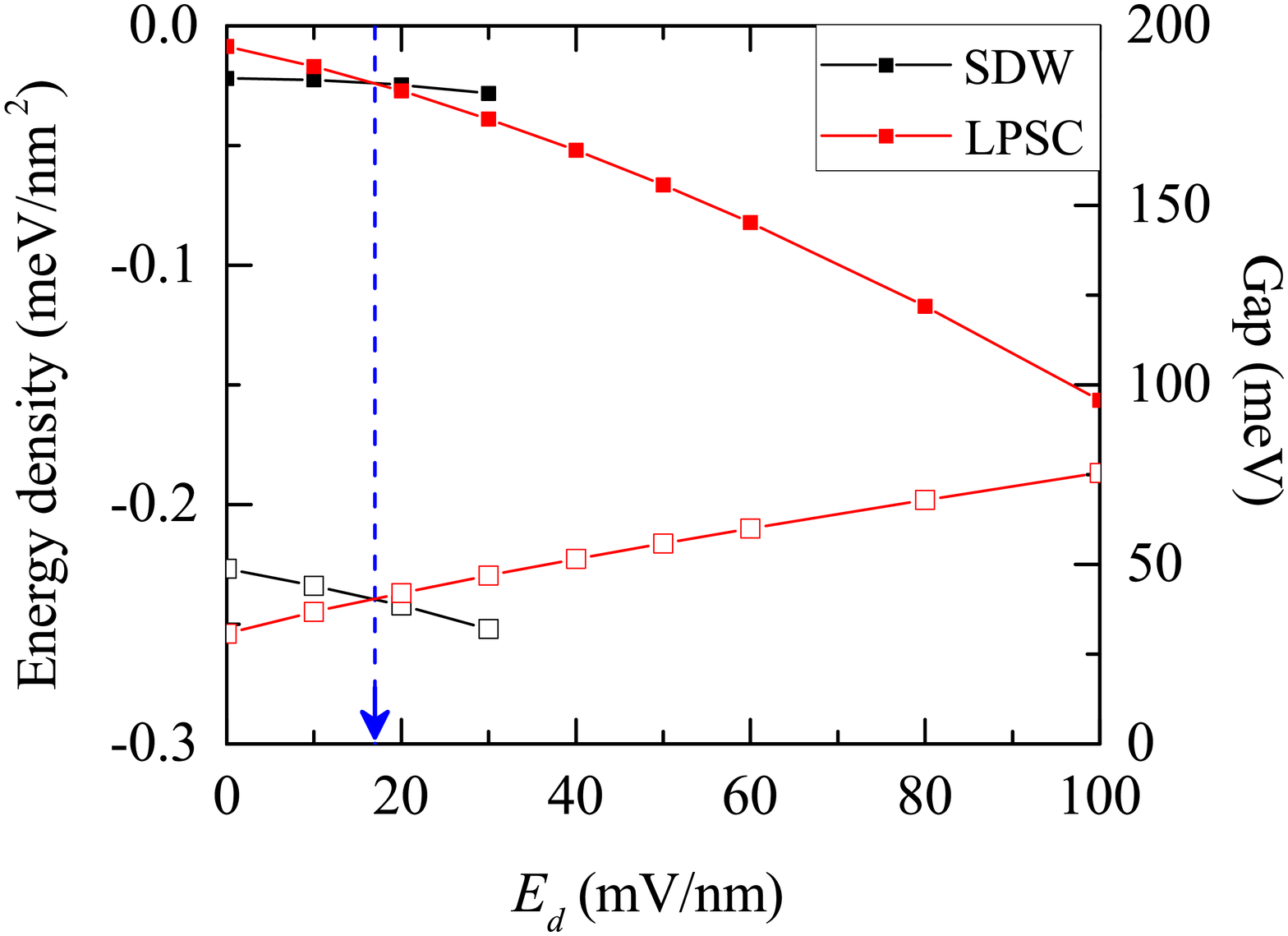}
\caption{Total energy and quasiparticle gap vs. applied electric field $E_d$. The black and red closed (open) squares represent the total energy (quasiparticle gap) for SDW and LPSC states respectively.  The SDW state is the ground state for small $E_d$ 
whereas the LPSC becomes the ground state for $E_d \gtrsim 17.0$ mV/nm, as indicated by the blue arrow. 
}
\label{fig:E&gap_1BL}
\end{figure}

The SDW state has no overall layer polarization at $E_d = 0$, as illustrated in Fig.~\ref{fig:E&gap_1BL},
and is the ground state \cite{MacDonald2012} because it avoids the electrostatic energy associated with spontaneous layer polarization.
In our calculations, which do not allow for spontaneous valley polarization, there is an first order phase 
transition between SDW and LPSC states at $E_d \sim 17$ mV/nm.  
In Fig.~\ref{fig:E&gap_1BL}, we plot the mean-field-theory quasiparticle energy gaps {\it vs.} $E_d$.
The mean-field-theory gap is $\sim$ 50 meV at $E_d=0$,
and increases up to 75 meV for $E_d \sim$ 100 mV/nm). 
The size of these gaps is strongly enhanced by the non-locality of exchange interactions,
as reflected also by the difference between LDA and GW approximation 
gaps in {\it ab initio} theories.\cite{BilayerGW} 
The difference is especially strong at small displacement fields,
since the gap vanishes at $E_d=0$ when exchange interactions are neglected.  
On the other hand, gaps are overestimated when exchange interactions are not screened.  

\section{Spatially Indirect Exciton Condensates in Double Bilayer Graphene:  $U_{b}=0$}

\begin{figure}[ht]
\includegraphics[width=.9\linewidth]{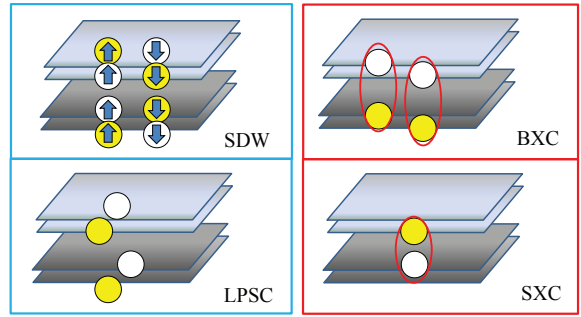}
\caption{Cartoon illustration of four double bilayer states, SDW, LPSC, SXC, BXC.
The red circles represent the dominant spontaneous coherence channel, as explained 
in the main text.}
\label{fig:Cartoon_2BL}
\end{figure}

We begin our exploration of the double bilayer phase diagram by focusing first on the $U_{b}=0$ line,  
first for the case of bilayer separation $t_{\rm hBN}=$ {0.3} nm
corresponding to one layer of hexagonal boron-nitride between graphene bilayers.
{For the double bilayer case we discuss here and later, we use the same set of screening parameters ($\epsilon=4$ and $C_s=0.8$) as that in the single bilayer calculation. Although the presence of one bilayer might further screen the other, this additional effect is limited as long as the electron density in each layer is still small, i.e. in the BEC regime. We therefore do not expect the phase boundary calculated to be very strongly altered by additional screening effects. \cite{screening}.}

In our calculation, we find that for $E_d \lesssim 70$ mV/nm, the mean-field bilayer state is not altered by the proximity of a 
neighboring bilayer.  Beyond this value of $E_d$, coherence develops 
between bilayers, and charge is transferred from the high potential
energy bilayer to the low potential energy bilayer.  
The critical value of $E_d$ at which spatially indirect coherence 
first develops can be identified with the displacement field at 
which the potential drop between bilayers is large enough to tune the lowest energy inter-bilayer 
spatially indirect exciton energy to zero.  

\begin{figure}[htb]
\includegraphics[width=.85\linewidth]{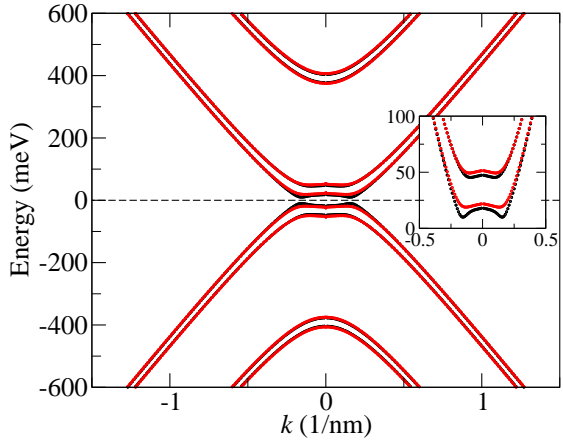}
\caption{Quasiparticle dispersions of the  single-layer exciton condensate state (SXC, $J_{X}=2$) and  the 
bilayer exciton condensate state (BXC, $J_{X}=0$).  
near the boundary between the LPSC and the BXC state at $E_d=80$ mV/nm and $U_{b}=0$.  
The BXC state is higher in energy because it is less able to polarize charge within each bilayer to 
take advantage of the displacement field.}
\label{fig:E_U=0.0_X(Ed=0.8)}
\end{figure}

\begin{figure}[htb]
\includegraphics[width=.95\linewidth]{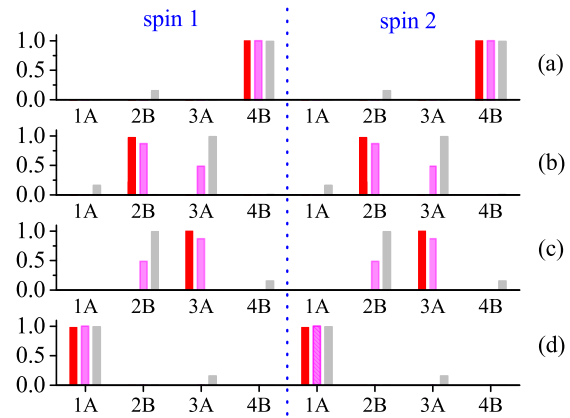}
\caption{Band projections of low energy $k=0$ eigenstates   
calculated for $E_d=80$ mV/nm and $U_{b}=0$.  
Panels (a), (b), (c), and (d) are for the two highest-energy valence
band states and the two lowest-energy conduction band states, ordered
by increasing eigenenergy.
The red, magenta, and gray solid bars represent site projections
calculated for the LPSC, SXC, and BXC states respectively.
Because none of these states break spin-symmetry, the quasiparticles energies
are doubly-degenerate and the orbital wavefunctions are spin-independent.}
\label{fig:Cmpnt_U=0p0_E=0.8.eps}
\end{figure}

Whenever isolated excitons have negative energies, their populations build up to finite 
values fixed by the repulsive\cite{SpatiallyIndirectExciton} exciton-exciton interaction strength, 
and they condense to yield 
spontaneous interlayer phase coherence.  As explained in Section II, the  
exciton condensate states of double bilayer graphene, and the corresponding isolated exciton states,
are distinguished by an interlayer chirality index, $J_{X}$.  The spontaneous coherence of the $J_{X}=2$ state is 
characterized by a relative phase between the 2B and 3A sites that is 
momentum-orientation independent.  This state maximizes the strong electron-hole interactions between
$\pi$-orbitals that are located on these adjacent layers, and we therefore refer to it as the 
single-layer excition condensate (SXC) state.  Because it produces electrons and holes unevenly within a 
bilayer, it is naturally layer polarized and takes good advantage of the displacement field within each bilayer
to lower its energy.  The $J_{X}=0$ state, on the other hand, has momentum-orientation independent inter-layer phases 
between bilayer valence band states shared between layers 1 and 2 and 
bilayer states shared between layers 3 and 4.  For this reason we refer to it as the 
bilayer exciton condensate (BXC) state.  The phase differences are momentum-orientation independent 
between layers 1 and 3 and between layers 2 and 4.  Because these layer-pairs are further apart than layers 2 and 3,
they have weaker electron-hole interactions.  Moreover the $J_{X}=0$ BXC state 
gains most inter-bilayer exchange energy when it is not layer polarized and is therefore less able to lower its energy in a 
displacement field.   On the other hand it gains more energy from hopping within the individual bilayers.  
Fig.~\ref{fig:Cartoon_2BL} contains a cartoon representation of the 
coherence patterns within the $J_{X}=0$ and $J_{X}=2$ states, and  
Figs.~\ref{fig:E_U=0.0_X(Ed=0.8)} and~\ref{fig:Cmpnt_U=0p0_E=0.8.eps} illustrate their
quasiparticle bands and coherence properties.   

The transition from the LPSC state to an excitonic condensate 
occurs at $E_d \sim 70$ mV/nm for the bilayer separation studied
and is from a state with no broken symmetries to a 
SXC state, rather than to a BXC state. 
We now explain the microscopic physics responsible for this choice.
The main panel in Fig.~\ref{fig:E_U=0.0_X(Ed=0.8)} compares the 
quasiparticle dispersions of the SXC and BXC states at $E_d = 80$ mV/nm, 
inside but close to the boundary of the exciton condensate portion of the phase diagram.    
(The eight bands plotted are all doubly degenerate because neither the BXC state nor the SXC state  
break spin-rotational invariance.)  We concentrate on the four doubly degenerate quasiparticle bands that are close to the 
Fermi energy (two above and two below.) 
For large momenta the bands are virtually identical in SXC and BXC states,
and approach those of two isolated bilayers with an energy offset 
equal to the electric potential difference $e (t+d) E_d$ between the two bilayer systems.
Differences between SXC and BXC states are found only in the small $k$ region highlighted 
in the inset, where we see that the band dispersions are flatter and that
the gap is slightly larger in the SXC state.  

More information on the low-energy eigenstates can be obtained by projecting them onto the 
site-localized basis set used for the Hartree-Fock calculations.  
Fig.~\ref{fig:Cmpnt_U=0p0_E=0.8.eps} plots $k=0$ eigenstates of the eight bands close to the 
Fermi level project onto A and B sublattice sites at $E_d=80$ mV/nm and $U_{b}=0$.
The projections onto 1B, 2A, 3B, and 4A are neglected because these sites have strong 
single-particle coupling to an adjacent layer and have high weight not in the bands close to
the Fermi level, but in the higher energy bands that are further from the Fermi level. 
Panels (a), (b), (c), (d) are in the order of ascending energy; the 
projections plotted in (a) and (b) are for occupied dressed valence bands
while those in (c) and (d) are for unoccupied conduction bands.
Because spin-rotational symmetry is not broken, the degenerate $\uparrow$ and $\downarrow$ 
bands in each panel have identical wavefunctions.  
The color codes identify the distinct many-body states: 
the red, magenta, and gray solid bars are $k=0$ projections for the 
LPSC, SXC, and BXC states respectively.

As seen in Fig.~\ref{fig:Cmpnt_U=0p0_E=0.8.eps}, the $k=0$ eigenstates of the LPSC, 
are completely localized in energetic order on sites 4B, 2B, 3A, and 1A.  
(The degree of layer polarization of the 
band eigenstates decreases as wavevector magnitude $k$ increases.)  
Site 2B is occupied before site 3A, even though layer 3 has a lower external potential 
than layer 2 because of the exchange-enhanced gap in an isolated bilayer discussed in 
Section III.   
The SXC state $k=0$ bands are similar to the LPSC bands in composition,
except that the bands closest to the Fermi energy, originally localized on 2B and 3A sites, 
hybridize.  In the BXC state the strongest hybridization occurs between sites whose 
layer indices differ by 2.   Note that layer 3 has higher weight at $k=0$ than layer 2 
in the occupied valence states in the BXC and that the 
condensation energy associated with excitons is reduced.  
For this reason the SXC state is always the ground state beyond the critical displacement field.  

\begin{figure}[htb]
\includegraphics[width=.95\linewidth]{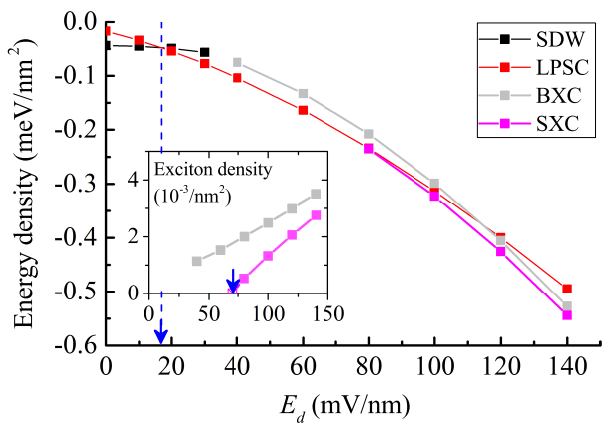}
\caption{Total energy vs. $E_d$ for SDW, LPSC, SXC, and BXC states. 
The inset plots the exciton densities of the SXC and BXC states.  
The SDW is the lowest energy state for small $E_d$. 
With increasing $E_d$ the ground state of the system 
first transforms into the LPSC state via a first-order phase transition 
and then into the SXC state via a continuous phase transition.
These results were obtained for hexagonal boron nitride barrier thickness
$t_{\rm hBN}= {0.3}$ nm.}
\label{fig:GEnDenvsEd_U=0p0}
\end{figure}    
 
The energetic comparison between LPSC, SXC and BXC states along the $U_b=0$ line is 
summarized in Fig.~\ref{fig:GEnDenvsEd_U=0p0}. 
The BXC state is metastable over a broad range $E_d$ values, but is never the ground state.
The SXC state, emerges from the LPSC via a continous phase transition.
As illustrated in the inset of Fig. \ref{fig:GEnDenvsEd_U=0p0}
the exciton density, which is identified as the electron density transferred from the 
high-electric-potential bilayer to the low-electric-potential bilayer, 
grows continuously from zero as $E_d$ increases beyond the value at which the 
LPSC becomes unstable.  

\begin{figure}[htb]
\includegraphics[width=.95\linewidth]{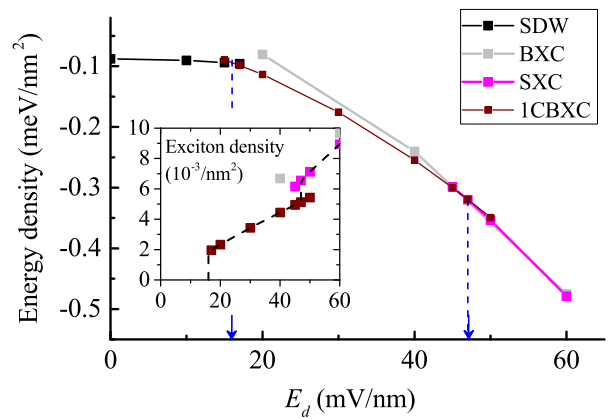}
\caption{Total energy vs. $E_d$ for SDW, BXC, SXC, and 1CBXC states. 
The inset plots the exciton densities of the SXC and BXC states.  
The SDW is the lowest energy state for small $E_d$. 
With increasing $E_d$ the ground state of the system 
first transforms into a LPSC via a first-order phase transition 
and then into the SXC state via a continuous phase transition.
These results were obtained for $t_{\rm hBN}= {0.9}$ nm}
\label{GEnDenvsEd_U_0p0_t_1p2}
\end{figure}

We now discuss the case of larger but still moderate bilayer separation.  
Total energy results for $t_{\rm hBN}= {0.9}$ nm,
corresponding to a three hBN layer barrier, 
are illustrated in Fig.~\ref{GEnDenvsEd_U_0p0_t_1p2}.
(We have also performed a similar calculation for $t=1.8_{\rm hBN}$ nm, corresponding to a six hBN-layer barrier,
which yielded results that are qualitatively similar except that phase boundaries are shifted to lower $E_d$.) 
The ground states in the 
small and large displacement field limits are SDW and SXC states, 
as in the small bilayer separation case.

{For a larger value of $t_{\rm hBN}$, a given displacement field yields a 
larger electric potential difference between the bilayers. 
As a result the spatially indirect band gap closes at smaller $E_d$ values, 
leading to condensate formation before the isolated bilayer SDW to LPSC transition 
occurs.  The ground state at intermediate displacement fields in this case 
is the 1CBXC state -- in which a spatially indrect exciton condensate forms for one spin species only.
We can view the effect of increasing $t_{\rm hBN}$ as equivalent to 
an increase in $U_b$ that is proportional to $E_d$, combined with a 
decrease in the inter-bilayer Coulomb interaction scale. 
The effective bias potential increase is $\pm e E_d \delta t_{\rm hBN}$ 
when the layer separation is increased by $\delta t_{\rm hBN}$. 
That effective bias potential addition favors charge transfer between bilayers and the formation of exciton
condensate states at smaller values of $E_d$.  
At the same time, weakening of inter-bilayer Coulomb interactions favors electron-hole pairs 
that form between the nearest layers in which the Coulomb interaction is maximzied.}  

\section{Spatially Indirect Exciton Condensates in Double Bilayer Graphene:  $E_{d}=0$}

\begin{figure}[htb]
\includegraphics[width=.95\linewidth]{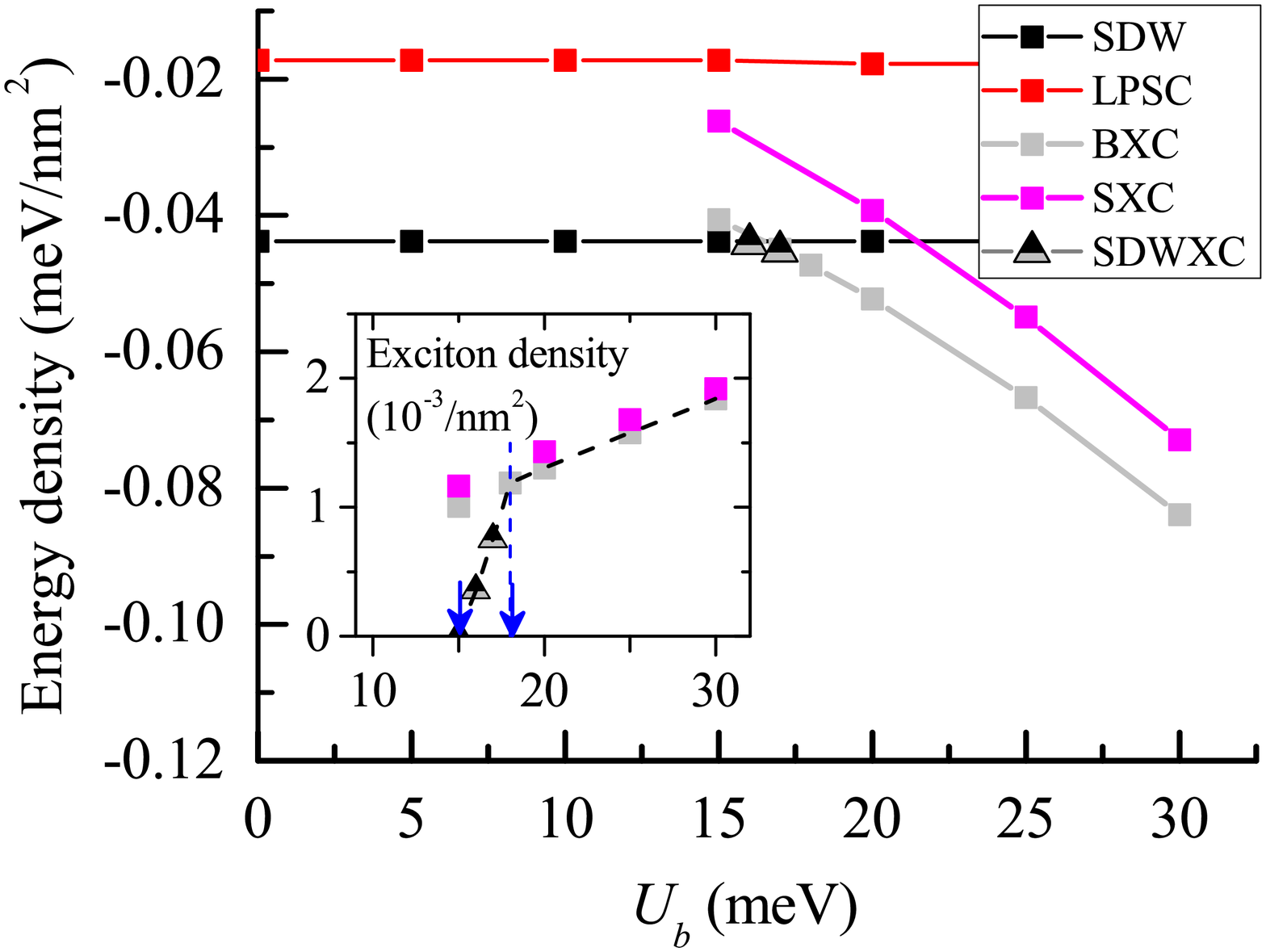}
\caption{Total energy vs. $U_b$ for SDW,LPSC,BXC,SXC and SDWXC states. 
The inset shows exciton densities in the three states with spatially indirect condensation.  
The SDW is the ground state for small $U_b$, and becomes unstable when its spatially indirect exciton energy 
vanishes.  The exciton condensate initially breaks spin-symmetry by condensing while 
retaining spin-density-wave order (SDWXC), but spin-symmetry is restored at higher exciton density. 
Both SDW to SDWXC, and SDWXC to BXC transitions are continuous.
}
\label{fig:GGEnvsU_Ed=0p0}
\end{figure}   

Next we study ground state as a function of bias voltage $U_b$
along the zero displacement field $E_d$ line. 
$U_b$ effectively shifts the relative energy of states in different 
bilayers without introducing a displacement field within the bilayers.
We have compared the total energies of possible ground states with different symmetries at each $U_b$,
as illustrated in Fig.~\ref{fig:GGEnvsU_Ed=0p0}.   
States can take advantage of $U_b$ only by shifting charge from bilayer to bilayer.
For this reason, the energies for both SDW and LPSC states are 
precisely constant as a function of $U_b$. 
The SDW state therefore remains the ground state until it becomes unstable
when its spatially indirect exciton energy vanishes.

We define the N{\' e}el order parameter vector of a bilayer SDW state as
$\vec{N}=\vec{s}_{l=1}-\vec{s}_{l=2}$, {\it i.e.} as the spin density difference between its 
top and bottom layers.  
As long as their is no single-particle 
tunneling between bilayers, the energy of the system is independent of the direction of either N{\' e}el 
vector.  To simplify the following discussion we assume that the two N{\' e}el vectors are antiparallel, in the 
$\hat{z}$-direction in the top $n=1$ biayer and in the $-\hat{z}$ direction in the bottom $n=2$ bilayer.
Then the $\uparrow$ valence band holes of the bilayer $n=1$ SDW state are concentrated 
in layer $l=1$ whereas the $\downarrow$ valence band holes are concentrated in layer $l=2$.  
Similarly for the lower $n=2$ bilayer the $\uparrow$ conduction band electrons are concentrated in layer $l=3$ 
whereas the $\downarrow$ conduction band electrons are concentrated in layer $l=4$.  
Because of the N{\' e}el order, the spatially indirect 
exciton energies are spin-dependent, with the lowest energy excitons having the 
same sense of layer polarization in each bilayer.  

\begin{figure}[ht]
\vspace{.8cm}
\includegraphics[width=.9\linewidth]{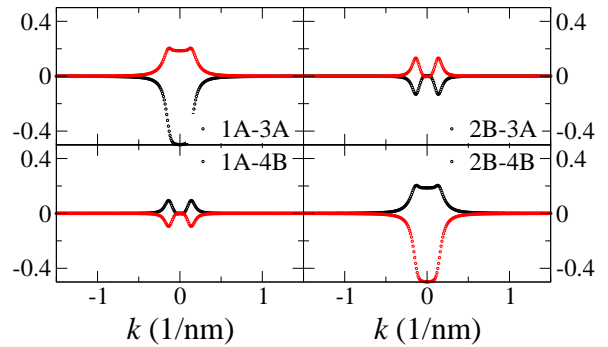}
\caption{Typical inter-sublattice coherence 
{\it vs.} wavevector $k$ for the SDWXC state when the N{\' e}el order 
parameter vectors in the two layers are along $\hat{z}$ and -$\hat{z}$ directions.   
The black and the red lines illustrate the coherence 
properties of $(\uparrow_e,\uparrow_h)$ 
and $(\downarrow_e,\downarrow_h)$ exciton condensates.}
\label{fig:coh_U=0p1_SDW1A3A(Ed=0p15)}
\end{figure}

\begin{figure}[htb]
\includegraphics[width=.95\linewidth]{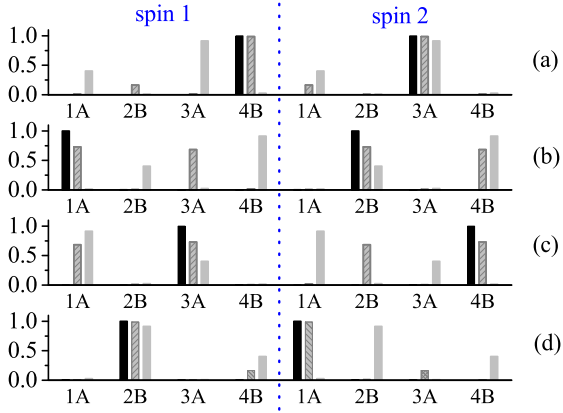}
\caption{Sublattice projections of SDW, SDWXC, and BXC $k=0$ quasiparticle wavefunctions
at $U_b=17$ meV and $E_d=0$ for the case in which the
N{\' e}el order parameter vectors in the two layers are along $\hat{z}$ and -$\hat{z}$ directions. 
The top four panels are for $\uparrow$ and $\downarrow$ valence band quasiparticles,
and the bottom four panels for $\uparrow$ and $\downarrow$ quasiparticles of conduction band quasiparticles.
The second and third rows report projections for the bands closest to the Fermi level as in  
Fig.\ref{fig:Cmpnt_U=0p0_E=0.8.eps}.  
The black solid, gray with slash pattern, and gray solid bars represent
sublattice projections for SDW, SDWXC, and BXC states respectively.}
\label{fig:Cmpnt_U=0.17_E=0.0}
\end{figure}

The character of the SDWXC state which forms along the $E_{d}=0$ line when the lowest exciton energy 
vanishes is illuminated by the coherence profile in
Fig.~\ref{fig:coh_U=0p1_SDW1A3A(Ed=0p15)}.
Because spin-invariance is still broken, the coherence properties are spin-dependent.  
For the N{\' e}el vector
directions we have chosen, it follows from the discussion in the previous 
paragraph that the lowest energy excitons are formed 
between electrons whose spins are parallel in the two bilayers;
$(\uparrow_e,\uparrow_h)$ excitons have dominant coherence between layers 1 and 3 while 
$(\downarrow_e,\downarrow_h)$ excitons have dominant coherence between layers 2 and 4.
(Note that positive $U_{b}$ favors 
holes in layers $l=1,2$ (bilayer $n=1$) and electrons in layers $l=3,4$ (bilayer $n=2$).) 
The weaker coherence between layers 2 and 4 for $(\uparrow_e,\uparrow_h)$ excitons and 
between layers 1 and 3 for $(\downarrow_e,\downarrow_h)$ excitons has a sign change
relative to the dominant coherence because of the spin and layer dependent mean-field potential responsible 
for gaps in the SDW state has opposite sign in the two layers.  
This picture of the SDWXC state is reinforced in Fig.~\ref{fig:Cmpnt_U=0.17_E=0.0}),
where sublattice projections are compared with those of the competing BXC state that has the same value of 
$J_{X}$ but does not break spin-rotational invariance.  The phase transitions between the SDWXC intermediate 
state and the small $U_{b}$ SDW state and the large $U_{b}$ BXC state are both continuous.  
The first phase transition adds spontaneous interlayer coherence to the previously established 
broken time-reversal symmetry, and the second phase transition drops broken
time reversal symmetry while maintaining spontaneous interlayer coherence.    

Because spatially indirect excitons first condense when their total excitation energy vanishes,
exciton binding energies ({\it i.e.} exciton energies relative to band gaps)
can be extracted from our calculations whenever the transition to a condensed state 
is continuous.  For example at $E_{d}=0$, the quasiparticle gap of the SDW state can be 
read off Fig. \ref{fig:E=0.0_1BL} and is $\sim 50$ meV.  From Fig.~\ref{fig:GGEnvsU_Ed=0p0}),
exciton condensation occurs at $U_b=15$ meV.   It follows that the exciton binding energy is 
50-15=35 (meV).  In addition to the exciton binding energy, our calculations also provide an 
estimate of the exciton-exciton interaction strength at low exciton densities, which can be obtained from
calculations of exciton density as a function of $U_b$.  
From the inset in Fig.~\ref{fig:GGEnvsU_Ed=0p0}, for example, we find that the interaction 
between the spatially indirect excitons of the SDW state  is repulsive with strength 
$\sim 17 {\rm eV nm}^{2}$.

\section{Spatially Indirect Exciton Condensates in Double Bilayer Graphene:  General Case}

In this section we briefly discuss some aspects of the phase diagram in Fig.~\ref{fig:Setup&PD} which 
do not emerge clearly from studies of the $U_{b}=0$ and $E_{d}=0$ lines.  
The SDW and LPSC, states, which do not have spontaneous interlayer phase coherence of the XC states, are
stable in the small-$E_{b}$, small-$U_{b}$ corner of the phase diagram.  These states are 
stable when the electrically controlled shift in the spatially indirect band gap is not large enough to reduce the smallest spatially indirect 
exciton energy to zero.  There is no mechanism to allow charge transfer between bilayers in the SDW and LPSC states.  
XC states appear at larger $U_{b}$ and $E_{b}$, because these electrical knobs both favor states that provide 
a mechanism for charge transfer.  
The broken symmetry SDW state is favored at small $E_d$ and the LPSC at larger $E_d$, 
as explained in Section III.  The spatially indirect XC states appear at larger $E_{d}$ and/or 
$U_{b}$, with large $E_d$ favoring SXC states and large $U_{b}$ favoring BXC states. 
The phase boundary between SXC and BXC states is first order, because the two states 
are distinguished by an integer-valued topological index $J_{X}$.  In addition to the SDWXC intermediate 
state, whose stability region includes a portion of the $E_{d}=0$ line, we have 
identified another distinct intermediate state with a stability region that does not 
include either the $E_{d}=0$ line or the $U_{b}=0$ line.  
For sufficiently large $E_{d}$, the transition from SDW to BXC states
with increasing $U_b$ occurs not via an intermediate SDWXC state but via a
state we refer to as a 1CBXC state, in which coherence is established only for one spin-component.
The 1CBXC state is characterized more fully below.  
All phase boundaries of the 1CBXC state mark first order phase transitions.

\begin{figure}[htb]
\includegraphics[width=.95\linewidth]{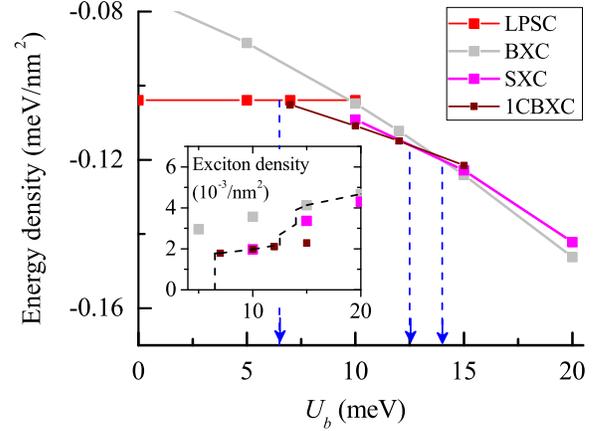}
\caption{Total energy vs. $U_b$ for various many body states at $E_d=40$ mV/nm.
The LPSC is lowest in energy state for small $U_b$. 
With increasing $E_d$ the ground state of the system first turns into an intermediate 1CBXC state 
with broken spin symmetry and coherence, and then into a BXC state.  
The inset shows jumps in exciton density at the LPSC to 1CBXC transition and 
at the 1CBXC to BXC state transtion, demonstrating that both transitions are first order.
}
\label{fig:GEnDenvsU_Ed=0p4}
\end{figure}

\begin{figure}[htb]
\includegraphics[width=.95\linewidth]{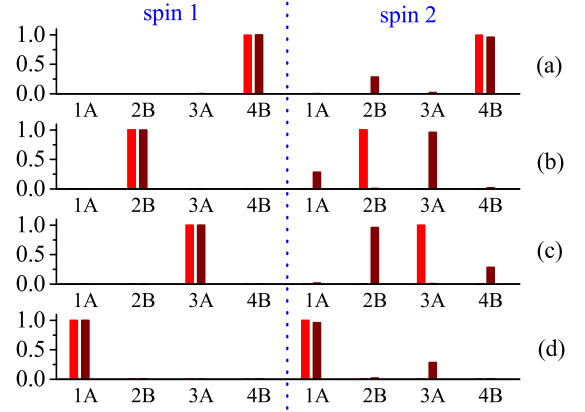}
\caption{Site projections of the low energy $k=0$ quasiparticle wavefunctions 
for $U_b=5$ meV and $E_d=40$ mV/nm, close to the LPSC to 1CBXC phase boundary.  
The red, brown, and gray solid bars represent the site quasiparticle wavefunction
site projections from the LPSC, 1CBXC, and BXC states respectively.}
\label{fig:Cmpnt_U=0.05_E=0.4}
\end{figure}   

\begin{figure}[htb]
\includegraphics[width=.95\linewidth]{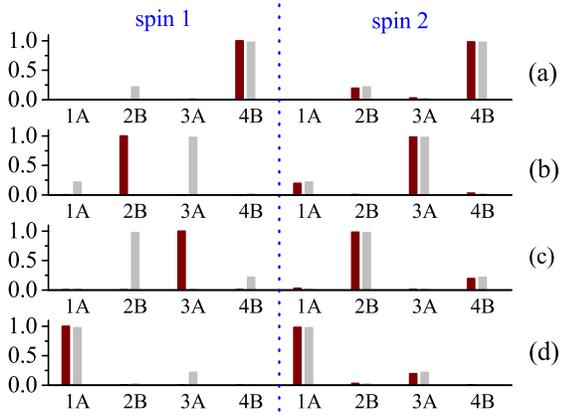}
\caption{Site projections of the low energy $k=0$ quasiparticle wavefunctions 
for $U_b=15$ meV and $E_d=40$ mV/nm, close to the LPSC to 1CBXC phase boundary.  
The brown, gray, and magenta bars represent the quasiparticle wavefunction
site projections from the 1CBXC, BXC and SXC states respectively.}
\label{fig:Cmpnt_U=0.15_E=0.4}
\end{figure}   

To shed further light on the competitions between these states we study the $E_d = 40$mV/nm line (line (A) in 
Fig.\ref{fig:Setup&PD}) in detail. The dependence of the ground state energy on $U_b$ is plotted 
for competing states in Fig.~\ref{fig:GEnDenvsU_Ed=0p4}. 
As $U_b$ increases the ground state evolves from a LPSU at the smallest values of 
$U_b$, to a 1CBXC mixed state at $U_b \sim 7$ meV, to a 
SXC state at $U_b \sim 12.5$ meV, and finally into a BXC state at $U_b \sim 14$ meV. 
Transitions are rarely intuitive if they occur when both $E_d$ and $U_b$ are finite. 
To compare the states before and after the transition, we define an index $S(MS1, MS2)$ which 
characterizes the similarity of two many body states, $MS1$ and $MS2$. 
This index is computing by adding contributions from the eight $k=0$ eigenstates $ES$ that are closest to the 
Fermi surface: $S(MS1, MS2)= ({1}/{8}) \sum_{ES=1}^{8} S_{ES}$ with
\begin{eqnarray}
S_{ES}= \sum_{b} \langle ES; MS1 | \, b \rangle.
\langle  b \, | ES; MS2 \rangle  
\end{eqnarray}
Here $b$ is the sublattice index. $S_{ES}$ is the inner product between $| ES; MS2 \rangle $ and $MS2$ at $k=0$.
The index summarizes the information contained in the wavefunction projection 
diagrams in a convenient way.  For continuous phase transitions $S(MS1, MS2)$ approaches one when 
the states are examined close to the transition.  For first order transitions, the index can differ substantially from one. 

We now characterize the LPSC to 1CBXC transition by examining
the $S$-index and the projection diagrams of Fig.~\ref{fig:Cmpnt_U=0.05_E=0.4} and Fig.~\ref{fig:Cmpnt_U=0.15_E=0.4}. 
These figures follow the same scheme as all previous bar diagrams; the red, brown, gray solid bars represent the LPSC, 
the mixed 1CBXC state, and the BXC, respectively. 
These figure show first of all that one spin component of the
the 1CBXC state (brown) is non-excitonic and strongly layer polarized.  
At $U_b=5$meV the spin 1 $k=0$ eigenstate components (Fig.~\ref{fig:Cmpnt_U=0.05_E=0.4}) are 
identical for LPSC and the 1CBXC states. For spin 2, the highest two (panel (d)) and the lowest two (panel (a)) of the eight eigenstates 
are quite similar ($S_{ES} \sim 1$), whereas the two eigenstates closest to the
Fermi surface (panel (b) and (c)) show drastic differences ($S_{ES} \sim 0$). 
More specifically, we see that the 2B component of the 
LPSC transforms to 3A (and partly 1A) components at the transition to the 1CBXC. 
This difference implies charge transfer between bilayers. 
The small $S_{ES}$ from these two eigenstates corresponds to an abrupt increase in exciton population at the transition, 
as confirmed by the inset of Fig.~\ref{fig:GEnDenvsU_Ed=0p4}. 
The transition between the LPSC and the 1CBXC is first-order.
Increasing $U_b$ further, we hit a second transition point. 
The corresponding bar diagram (Fig.~\ref{fig:Cmpnt_U=0.15_E=0.4}) shows the $k=0$ 
eigenstate projections for 1CBXC (brown) and BXC (gray) states.
The transition from the 1CBXC state to the BXC  occurs mainly on spin 1 which undergoes a
transformation that is similar to that experienced by spin 2 during the LPSC to 1CBXC transition. 
A similar discontinuity in exciton density profile shows that hte 1CBXC to BXC 
transition is also a first-order phase transition. 
Transitions along this line have $S \sim 6/8 = 0.75$,  implying that 
six out of eight eigenstates are very similar between before and after the transition.        

\begin{figure}[htb]
\includegraphics[width=.95\linewidth]{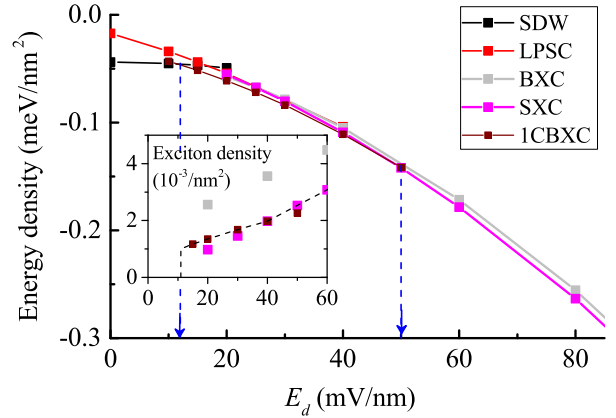}
\caption{Total energy vs. $E_d$ for various many body states at $U_b=10$ meV; the inset shows the corresponding exciton densities.
The SDW state has lowest energy for small $E_d$. 
With increasing $E_d$ the ground state transitions first into the mixed 1CBXC state
and then into the SXC state.}
\label{fig:GEnDenvsEd_U=0p1}
\end{figure}            

\begin{figure}[htb]
\includegraphics[width=.95\linewidth]{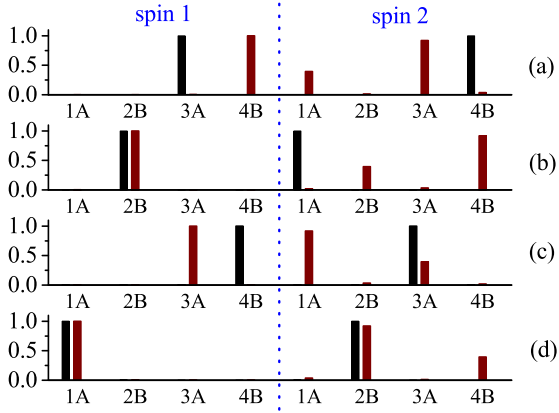}
\caption{Site projections of low energy $k=0$ quasiparticle wavefunctions 
for $U_b=10$ meV and $E_d=15$ mV/nm. The relationship between panel position and 
eigenenergy order is the same as in Fig.\ref{fig:Cmpnt_U=0p0_E=0.8.eps}.
The black, brown, and magenta solid bars characterize the SDW state, the 1CBXC state, and the SXC state respectively.  
The N{\' e}el state order parameter vectors of the two bilayers
are oppositely oriented.}  
\label{fig:Cmpnt_U=0.1_E=0.15}
\end{figure}  

\begin{figure}[htb]
\includegraphics[width=.95\linewidth]{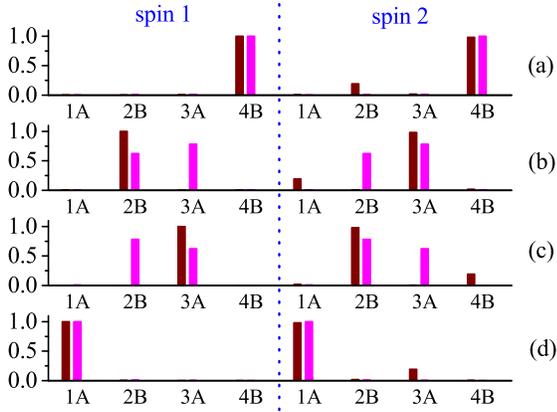}
\caption{Site projections of low energy $k=0$ eigenstates for $U_b=10$ meV and $E_d=50$ mV/nm, near
the phase transition between 1CBXC and SXC states.  
The brown and magenta bars represent the 1CBXC and the SXC states respectively.}
\label{fig:Cmpnt_U=0.1_E=0.5}
\end{figure}   

Next we fix the bias potential  at $U_b = 10$ mV and sweep $E_d$ (line (B) in Fig.~\ref{fig:Setup&PD}).  
The energy comparison in Fig.(\ref{fig:GEnDenvsEd_U=0p1})  shows that at low $E_d$, the ground state is the SDW. 
Increasing $E_d$ drives a first-order transition into the 1CBXC intermediate state at $E_d \sim 12$ mV/nm. 
This is followed by another transition into the SXC state at $E_d \sim 50$ eV/nm. We see from Fig.~\ref{fig:Cmpnt_U=0.1_E=0.15} that the transition from SDW to 1CBXC is first order and accompanied by a considerable charge redistribution;
only the eigenstates in panels (b), (d) of spin 1 and (d) of spin 2 have $S_{ES} \sim 1$.
The charge redistribution happens both within bilayers (panel (a) and (c) of spin 1, panel (a) of spin 2) as well as between bilayers (panel (b) and (c) of spin 2), which shows that electrons move from layer 1 to 4  (holes from layer 3 to 1). 
The significant change results in a low $S$-index of less than $3.5/8= 0.4375$. 
A transition between 1CBXC and SXC occurs then occurs at larger $E_d$.
By the time this phase boundary is reached 
the 1CBXC state has evolved into a state that is
similar to the SXC state so that the transition, although first order 
is characterized by a value of the similarity index $S$ that can be as large as $(4+4/\sqrt{2})/8 \sim 0.85$. 
Because of the greater similarity, the transition between 1CBXC and SXC, although is first order, 
appears to be smooth in the exciton density profile illustrated in the inset of Fig.(\ref{fig:GEnDenvsEd_U=0p1}).

\begin{figure}
\includegraphics[width=1.\linewidth]{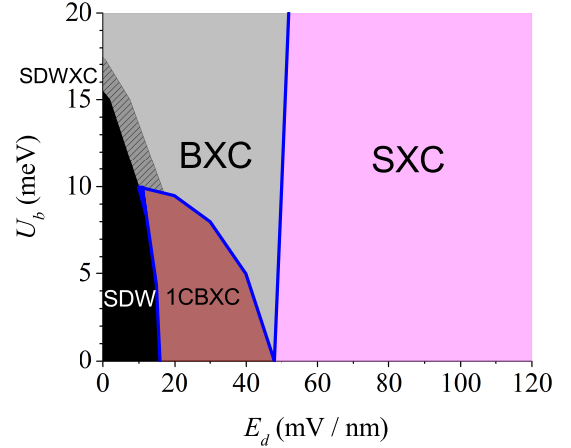}
\caption{Phase diagram for bilayer separation $t_{\rm hbN}= {0.9}$ nm.  The 
LPSC states does not appear because the spatially indirect exciton energy vanishes 
before the isolated bilayer SDW to LPSC transition occurs.  
The solid blue lines mark first order phase boundaries.} 
\label{fig:Gphasediagram_t=1p2}
\end{figure}  

So far we have discussed mainly results are for the case of a single-layer hBN barrier 
which has the richest phase diagram. 
We now explain how the topology of the phase diagram is altered by increasing the bilayer separation. 
Fig.~\ref{fig:Gphasediagram_t=1p2} shows the calculated phase diagram for bilayer separation $t_{\rm hBN}= {0.9}$ nm, 
corresponding to three-layer hBN barriers.  We first note that 
for $E_d<15$ mV/nm differences compared to the single-layer barrier are small.  
For larger $E_d$, the main change is that the LPSC state is absent in the three-layer barrier 
case.  Additionally, other phases including the 1CBXC and BXC states are 
shifted downward toward smaller $U_b$ compared 
to the single-layer phase diagram in Fig.~\ref{fig:Gphasediagram_t=1p2}. 
We can understand these differences by taking into considering the two main effects of increasing layer 
separation. First of all increasing $t_{\rm hBN}$ by $\delta t_{\rm hBN}$ increases the effective bias potential by $eE_d \delta t_{\rm hBN}$ 
because of the 
additional electric potential difference between bilayers at a given value of $E_d$.  
This simple shift accounts for the most of the barrier thickness dependence of the phase diagram.
The small $E_d$ SDW region of the phase diagram  is therefore relatively independent of $t_{\rm hBN}$.
The intermediate state 1CBXC on the other hand shifts noticeably toward smaller $U_b$ and squeezes out the LPSC. 
According to our calculations, the LPSC state is already entirely eliminated for the three-layer hBN barrier.  
In addition increasing the potential difference between bilayers, increasing $t_{\rm hBN}$ decreases the 
strength of inter-bilayer electron-electron interactions.
This second effect favors SXC states over BXC states because the relative reduction in interaction strength 
is larger for the adjacent layers of the two bilayer systems than for the more remote layers.
The phase boundary between BXC and SXC states therefore shifts toward smaller $E_d$, expanding the stability 
range of the SCX state.   
We also note that at still larger $t_{\rm hBN}$ separations, substantial charge transfer between 
bilayers will occur even at small $E_d$, very quickly driving the double-bilayer system 
into a metallic Fermi liquid state.

\section{Summary and Discussion}

In this paper we have systematically constructed a mean-field-theory phase diagram
for neutral double-bilayer graphene, demonstrating that the many-body ground state 
can be altered by an external displacement field $E_{d}$ and an interlayer bias $U_{b}$.
Mean-field theory predicts that the ground state of neutral isolated bilayers is a spin-density-wave
state at $E_{d}=0$ and a layer-polarized semiconductor at large $E_{d}$.  These properties 
have already been verified experimentally.  Both $E_{d}$ and $U_{b}$ favor 
charge transfer between bilayers, and this is accomplished in mean-field theory 
by forming excitonic condensate states that have a gap for charged excitations and 
spontaneous coherence between bilayers.  The onset of exciton condensation occurs when
the electrically tunable indirect exciton excitation energy becomes negative.
The layer, sublattice, and spin degrees of freedom of graphene bilayers allow for a variety of different single exciton states, and correspondingly for a 
variety of different condensate states.  Distinct excitonic states can be classified by a topological quantum number $J_{x}$,
related to relative angular momentum.   
We find that the combination of the possibility of broken time-reversal symmetry in spin-density-wave 
states and the possibility of condensation in different excitonic states, leads to the complex phase diagram presented in Fig.[1]. The phase diagram contains two types of intermediate states that combine broken spin-rotational invariance and exciton condensation in different ways.  

\begin{figure}
\includegraphics[width=1.\linewidth]{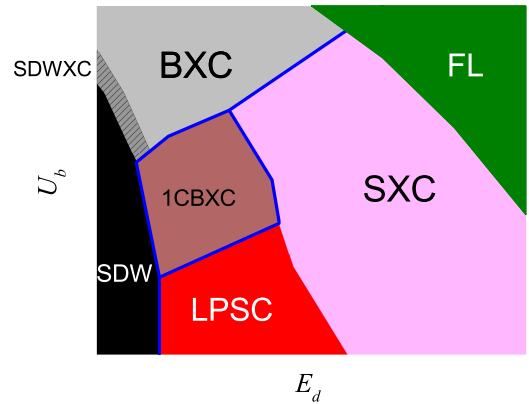}
\caption{Schematic phase diagram including Fermi liquid (FL) states.  
The mean-field theory instability of Fermi liquid states toward states 
with spontaneous coherence between bilayers is expected to be 
suppressed by correlations at high exciton densities.}
\label{fig:Gphasediagram_FL}
\end{figure}

The mean-field theory we use in this paper, which is closely related to the BCS mean-field theory of 
superconducting states, can capture the physics of excitons in the BEC regime and the crossover to the 
high exciton-density BCS regime.  It does not however account for the 
metallic Fermi liquid (FL) phase generally expected\cite{Neilson2013, Maezono_PRL2013} at high exciton 
densities.  The FL phase, like the exciton condensate, has charge transfer between bilayers,
but does not support broken symmetries of charge-excitation gaps.  
Because these simple Fermi liquid states are not predicted by mean-field theory,
in which condensation always occurs when excitons are present, we have chosen not to represent them
explicitly in Fig.[1].  Metallic Fermi liquid double-bilayer excitonic states are expected when 
the density of excitons times the area occupied by a spatially indirect exciton bound state is large,
in other words when excitons overlap strongly, and therefore should appear along the upper right 
of the phase diagram in Fig.[1] as indicated schematically in Fig.[~\ref{fig:Gphasediagram_FL}]. 
Another possible phase that is not described by the mean-field theory is the biexciton phase\cite{Maezono_PRL2013} in which 
excitons pair to form bound biexcitons which are analogs of hydrogen molecules. In mean-field theory, the interactions between excitons is always repulsive.
The attractive interaction that can result from fluctuating dipoles can in principle produce attraction.
This effect is unlikely to be important at finite $E_{d}$ however, because the dipole orientations are fixed by 
displacement fields.

The excitonic condensates proposed in this paper are normally most conveniently 
identified in experiment by performing drag measurements, since spontaneous coherence in an exciton state 
shorts electrical isolation and eliminates layer dependence of measured voltages, even 
when separate contacting is well established experimentally.  In a double-bilayer drag experiment current 
flows through one bilayer and voltage drops are measured both in the bilayer carrying the 
transport current (the drive layer) and in the adjacent electrically isolated bilayer (the drag layer). 
Consider for example the $E_{d}=0$ case.  At $U_{b}=0$ both drag and drive layers are expected to be 
in SDW states if disorder is sufficiently weak.  The resistance measured in an isolated neutral bilayer 
in an SDW state should therefore be large and increase indefinitely as temperature is lowered.  
(Importantly, this property contrasts with the extensively studied spatially-indirect-condensate case 
of semiconductor double quantum well systems in a 
magnetic field, where coherence is established between layers
in $\nu=1/2$ quantum Hall states.  Although the quantum Hall spatially indirect 
condensate state has a bulk gap, it also has topologically protected edge states 
and therefore has a finite longitudinal resistance as temperature goes to zero.)
By increasing $U_{b}$ the spatially indirect exciton energies can be tuned to zero, allowing 
excitons to be present in the system in equilibrium.  We predict that when the exciton density is low,
spontaneous interlayer coherence will be established between the bilayers and the resistive 
voltage drop measured in the direction of current flow will be detected by voltage probes 
connected to either bilayer.  

We emphasize that there is a qualitative difference between double bilayer graphene and double monolayer 
graphene\cite{Min_PRB2008, Lozovik_JETPLett2008, Bistritzer_PRL2008, Kharitonov_PRB2008, Bistritzer2010,Mink2011,Lozovik_PRB2012, Fischetti2014}
that is relevant to their ability to host robust spatially indirect exciton condensates.
Because the bilayer is a semiconductor, it has a band of elementary neutral excitations, the exciton states,
that lie below the particle-hole excitation spectrum continuum.  
As long as the density 
of excitons is low, the argument that they can be considered to be weakly interacting bosons is 
straightforward and reliable.  The weakly-interacting low-exciton-density limit is accurately captured by the mean-field
approximation that we employ, whereas quantum fluctuations become more important at higher 
exciton densities.  It is generally expected that the true ground state at high exciton densities is a 
Fermi liquid, as discussed in the previous paragraph, although quantitative calculations remain a challenge.
At a minimum the critical temperature in this limit is expected to be reduced because electron-hole 
interactions are screened.\cite{Kharitonov_PRB2008,Bistritzer2010,Fischetti2014, Neilson_PRB2014}
Because monolayer graphene is a gapless semiconductor, its particle-hole excitation continuum does not 
have a lower bound and there are therefore no isolated bosonic excitations.  Similarly, in double monolayer 
graphene there are no isolated spatially-indirect exciton excited states\cite{Lozovik_PRB2012}.  The low-density regime 
in which more quantitative theoretical predictions are possible is absent in the double-monolayer case. 
{Because of its electrically tunable gaps, the double bilayer provides an interesting opportunity to explore the crossover between the rather simple case of spatially indirect condensates formed between two-dimensional semiconductors, 
and the more complex case of spatially indirect exciton condensates in gapless systems, which is approached for small displacement fields. } 

Although this work has restricted its attention to the physics of neutral double-bilayers, it seems clear that
inter-bilayer electron-electron interactions can also have a strong influence on double-bilayer properties 
away from charge neutrality, particularly when the carrier density in one bilayer or the other is low and 
the Fermi level lies close to the Mexican hat features in the quasiparticle dispersion that are evident in the 
quasiparticle band dispersions plotted in this manuscript. 
{A number of recent experimental papers\cite{Li_arxiv2016(Dean_doublebilayer), Liu(Kim), Neilson2013,Zarenia_SciRep2014,Tutuc2016} 
with intriguing findings demonstrate the potential for interesting many-electron physics in the double-bilayer system,
a part of which is addressed here, leaving many avenues for future theoretical and experimental work.}

\section{Acknowledgments}
We thank  M. Allen, C. Dean, P. Kim, C. N. Lau, J. Li, X. Liu, F. Wang, A. Yacoby and A. Young for valuable interactions.
J-JS acknowledges the supports by Ministry of Science and Technology (MOST 102-2112-M-009-018-MY3) and by 
National Center for Theoretical Sciences of Taiwan.  AHM acknowledges the supports from the NRI SWAN program and 
from the Welch Foundation grant TBF1473.


\begin{thebibliography}{}
\bibitem{CommentAHE}  We neglect the possibility of broken valley symmetries.  When this type of 
broken symmetry is allowed, it is possible that intermediate  anomalous 
quantum Hall states occurs over a range of displacement fields.
\end{thebibliography}
\end{document}